\documentclass[twocolumn,a4,showpacs,preprintnumbers,amsmath,amssymb]{revtex4-1}
\usepackage{longtable}
\usepackage{epsfig}

\usepackage{graphicx}
\usepackage{dcolumn}
\usepackage{bm}
\usepackage{mathptmx, courier, pifont}
\usepackage[scaled=1.0]{helvet}
\usepackage[T1]{fontenc}
\usepackage{textcomp}
\usepackage{color}
\usepackage{colordvi}

\begin{document}

\title{ Nature of $\gamma$-deformation in Ge and Se nuclei and the
triaxial projected shell model description }

\author{G. H.~Bhat$^{1}$,  W. A.~Dar$^{1}$, J. A.~Sheikh$^{1,2}$
and Y.~Sun$^{3,4,2}$\footnote{Corresponding author at SJTU:
sunyang@sjtu.edu.cn}}

\address{$^1$Department of Physics, University of Kashmir, Srinagar
190 006, India \\
$^2$Department of Physics and Astronomy, University of
Tennessee, Knoxville, TN 37996, USA \\
$^3$Department of Physics and Astronomy, Shanghai Jiao Tong
University, Shanghai 200240, People's Republic of China \\
$^4$Institute of Modern Physics, Chinese Academy of Sciences,
Lanzhou 730000, People's Republic of China }

\date{\today}

\begin{abstract}
Recent experimental data have demonstrated that $^{76}$Ge may be a
rare example of a nucleus exhibiting rigid $\gamma$-deformation in
the low-spin regime. In the present work, the experimental analysis
is supported by microscopic calculations using the
multi-quasiparticle triaxial projected shell model (TPSM) approach.
It is shown that to best describe the data of both yrast and
$\gamma$-vibrational bands in $^{76}$Ge, a rigid-triaxial
deformation parameter $\gamma\approx 30^\circ$ is required. TPSM
calculations are discussed in conjunction with the experimental
observations and also with the published results from the spherical
shell model. The occurrence of a $\gamma\gamma$-band in $^{76}$Ge is
predicted with the bandhead at an excitation energy of $ \sim$ 2.5
MeV. We have also performed TPSM study for the neighboring Ge- and
Se-isotopes and the distinct $\gamma$-soft feature in these nuclei
is shown to result from configuration mixing of the ground-state
with multi-quasiparticle states.
\end{abstract}


\maketitle

\section{Introduction}

Atomic nuclei are among the most fascinating quantum many-body
systems that depict a rich variety of shapes and structures
\cite{BM75}. Most of the nuclei are known to have axially-symmetric,
dominantly quadrupole, deformed shape in the ground-state. However,
there are also regions in the nuclear periodic table, referred to as
transitional regions, where axial-symmetry is broken and a triaxial
mean-field description is appropriate to characterize the properties
of these nuclei \cite{JS99}. For nuclei depicting triaxial shapes,
there is a long-standing issue whether these nuclei have rigid or
soft $\gamma$-deformation (see, for example, discussions in Refs.
\cite{SW01,GS03,DJ09}). Traditionally, there are two extreme
phenomenological models that describe the triaxiality: the one with
a rigid-$\gamma$ deformation of Davydov and Flippov (DF) \cite{AS58}
and the $\gamma$-soft model of Wilets and Jean \cite{LM56}. Both
models give rise to similar level energies and $B(E2)$ transition
strengths for the ground-state bands and, therefore, it is
impossible to delineate the two different modes of excitations. In
fact, there have been suggestions \cite{OS87} that the two
descriptions are equivalent and intrinsic single-particle wave
functions obtained from a triaxially-deformed well are useful in
describing low-lying collective states. However, it has been
demonstrated \cite{NV91} that the phase of the odd-even staggering
(i.e. the staggering of the odd- and even-spin levels) of the
observed $\gamma$-bands could shed light on the nature of the
triaxial shape with rigid-$\gamma$ rotation exhibiting an opposite
staggering pattern to that of the $\gamma$-soft case.

Recently, using the experimental techniques of above-barrier Coulomb
excitation and inelastic scattering, $\gamma$-band energies of
$^{76}$Ge have been extended considerably \cite{YT13}. It has been
shown that the odd-even staggering of the $\gamma$-band is quite
opposite to that of all neighboring nuclei and is in conformity with
that expected for a rigid-$\gamma$ deformation \cite{YT13,EA07}.
This is one of rare examples of atomic nuclei exhibiting
rigid-$\gamma$ deformation in the low-lying states. The observed
yrast- and excited states have been discussed \cite{YT13} using the
DF model and also the spherical shell model (SSM) approaches. In the
SSM approach \cite{NY08}, the pairing plus quadrupole-quadrupole
interaction was employed in the
$\{g_{9/2},\,p_{1/2},\,p_{3/2},\,f_{5/2}\}$ configuration space, and
it has been demonstrated that SSM provides a very good description
of the observed data for the low-lying states in $^{76}$Ge.

The purpose of the present work is to investigate the high-spin
properties of $^{76}$Ge using the multi-quasiparticle (qp) triaxial
projected shell model (TPSM) approach
\cite{GH08,JG09,JG11,GJ12,Ch12}. In the SSM analysis, the primary
emphasis was on the low-lying states and the present investigation
complements the results obtained by the SSM approach. In TPSM, apart
from 0-qp, 2- and 4-qp configurations are explicitly included in the
basis space. Therefore, in this model it is possible to investigate
the high-spin band-structures, which provides important information
on the interplay between collective and single-particle excitations,
and thus to probe single-particle structures in the neutron-rich
mass region. In the present study, we have also performed a detailed
study of the neighboring nuclei to investigate the nature of
$\gamma$-deformation in these nuclei in comparison to $^{76}$Ge.

The manuscript is organized as follows. In the next section, we
provide a few details of the TPSM model for completeness and further
details can be found in our earlier publications
\cite{GH08,JG09,JG11,GJ12,Ch12,YK00}. Section III is completely
devoted to the investigation of $^{76}$Ge and in section IV, the
results of the neighboring Ge- and Se-isotopes are presented and
discussed. Finally, in section V, we provide a summary of the work
performed in the present manuscript.

\section{Outline of the triaxial projected shell model}

In TPSM, triaxially-deformed Nilsson states are employed as a
starting basis to describe a nucleus exhibiting axial and triaxial
deformations. An explicit three-dimensional angular-momentum
projection is then performed for configurations built from the
deformed Nilsson states. A triaxial qp configuration is an admixture
of different $K$ (projection along the symmetry axis) states, and
the vacuum configuration is composed of $K=0,2,4,...$
states for an even-even system. It was shown \cite{YK00} that the
angular-momentum projection from the $K=0$, 2, and 4 states
correspond to the ground, $\gamma$- and $\gamma\gamma$- bands,
respectively. The model has recently been extended
\cite{GH08,JG09,JG11,GJ12,Ch12,GC06,SB10} to include multi-qp
configurations in the model space, which allows one to describe
states of collective $\gamma$-vibrations and qp excitations on an
equal footing. For instance, the multi-qp TPSM approach has been
used to investigate the interplay between the vibrational and the
quasi-particle excitation modes in $^{166-172}$Er \cite{JG11}. It
was demonstrated that a low-lying $K=3$ bands observed in these
nuclei, the nature of which had remained unresolved, are built on
triaxially-deformed 2-qp states. This band is observed to interact
with the $\gamma$-vibrational band and becomes favored at high
angular-momentum for some Er-nuclei. In another study \cite{Ch12},
the long-standing puzzle of the very different $E2$ decay rates from
the same 2-quasineutron $K^\pi = 6^+$ isomers in the $N = 104$
isotones was investigated. It was shown that the highly
$K$-forbidden transition from the $6^+$ isomer to the ground-state
band is sensitive to the mixing with the $6^+$ state of the
$\gamma$-vibrational band.

For even-even systems, the TPSM basis are composed of 0-qp (qp
vacuum), 2-proton, 2-neutron, and 4-qp configurations, i.e.,
\begin{eqnarray}
\{ \hat P^I_{MK}\left|\Phi\right>, ~\hat P^I_{MK}~a^\dagger_{p_1}
a^\dagger_{p_2} \left|\Phi\right>, ~\hat P^I_{MK}~a^\dagger_{n_1}
a^\dagger_{n_2} \left|\Phi\right>,  \nonumber \\~\hat
P^I_{MK}~a^\dagger_{p_1} a^\dagger_{p_2} a^\dagger_{n_1}
a^\dagger_{n_2} \left|\Phi\right> \}, \label{basis}
\end{eqnarray}
where $P^I_{MK}$ is the three-dimensional
angular-momentum-projection operator \cite{RS80} and
$\left|\Phi\right>$ in (\ref{basis}) represents the triaxial qp
vacuum state. The qp basis chosen in (\ref{basis}) is adequate to
describe high-spin states up to $I\sim 20\hbar$ for even-even
systems. In the present analysis we shall, therefore, restrict our
discussion to this spin regime. It is noted that for the case of
axial symmetry, the qp vacuum state has $K=0$ \cite{KY95}, whereas
in the present case of triaxial deformation, the vacuum state is a
superposition of all possible $K$-values. Rotational bands with the
triaxial basis states, Eq. (\ref{basis}), are obtained by specifying
different values for the $K$-quantum number in the angular-momentum
projection operator \cite{RS80}.

\begin{table}
\caption{The axial deformation parameter, $\epsilon$, and triaxial
deformation parameter, $\epsilon'$, employed in the calculation for
$^{70-80}$Ge and $^{76-82}$Se. The $\gamma$ deformation is related
to the above two parameters through $\gamma = \tan^{-1}
(\epsilon'/\epsilon)$. $\epsilon$ is related to the $\beta$
deformation through $\epsilon = 0.95\times\beta$.}
\begin{tabular}{c|ccccccccccc}
\hline            & $^{70}$Ge &$^{72}$Ge & $^{74}$Ge & $^{76}$Ge & $^{78}$Ge & $^{80}$Ge &$^{76}$Se & $^{78}$Se & $^{80}$Se & $^{82}$Se \\
\hline $\epsilon$ & 0.235   & 0.230      & 0.220      & 0.200      & 0.210    & 0.200 &  0.260 &0.256  & 0.220  & 0.180  \\
       $\epsilon'$& 0.145   & 0.150      & 0.155      & 0.160      & 0.150    & 0.145 &  0.155 & 0.150 & 0.130  &
       0.130 \\
       $\gamma$   & 31.68   & 33.11      & 35.17      & 38.66      &
       35.54 & 35.94 & 30.81 & 30.37 & 30.58 & 35.84
\\\hline
\end{tabular}\label{TableDeforPara}
\end{table}
\begin{table}
\caption{The QQ-force strengths $\chi$ in unit of $10^{-2}$ MeV for
$^{70-80}$Ge and $^{76-82}$Se isotopes.   }
\begin{tabular}{cccc}
\hline
            &    $\chi_{nn}$       &    $\chi_{pp}$  &  $\chi_{np}$  \\\hline
$^{70}$Ge   &    8.9649     &  7.9944    & 8.4658            \\
$^{72}$Ge   &    8.7473     &  7.5382    & 8.1202            \\
$^{74}$Ge   &    8.5451     &  7.1283    & 7.8046             \\
$^{76}$Ge   &    7.8172     &  6.3219    & 7.0299            \\
$^{78}$Ge   &    8.0674     &  6.3338    & 7.1482            \\
$^{80}$Ge   &    8.6937     &  6.6345    & 7.5947            \\
$^{76}$Se   &    7.8240     &  6.7959    & 7.2918             \\
$^{78}$Se   &    7.6688     &  6.4576    & 7.0372            \\
$^{80}$Se   &    7.7027     &  6.2968    & 6.9643            \\
$^{82}$Se   &    8.0623     &  6.4064    & 7.1868             \\
\hline
\end{tabular}\label{ChiValue}
\end{table}

As in the earlier projected shell model \cite{KY95} calculations, we
use the pairing plus quadrupole-quadrupole Hamiltonian
\begin{equation}
\hat H = \hat H_0 - {1 \over 2} \chi \sum_\mu \hat Q^\dagger_\mu
\hat Q^{}_\mu - G_M \hat P^\dagger \hat P - G_Q \sum_\mu \hat
P^\dagger_\mu\hat P^{}_\mu . \label{hamham}
\end{equation}
Here $\hat H_0$ is the spherical single-particle Hamiltonian which
contains a proper spin-orbit force described by the Nilsson
parameters \cite{Ni69}. The QQ-force strength $\chi$ is related to
the quadrupole deformation $\epsilon$ as a result of the
self-consistent HFB condition and the relation is given by
\cite{KY95}:
\begin{equation}
\chi_{\tau\tau'} =
{{{2\over3}\epsilon\hbar\omega_\tau\hbar\omega_{\tau'}}\over
{\hbar\omega_n\left<\hat Q_0\right>_n+\hbar\omega_p\left<\hat
Q_0\right>_p}},\label{chi}
\end{equation}
where $\omega_\tau = \omega_0 a_\tau$, with $\hbar\omega_0=41.4678
A^{-{1\over 3}}$ MeV, and the isospin-dependence factor $a_\tau$ is
defined as
\begin{equation}
a_\tau = \left[ 1 \pm {{N-Z}\over A}\right]^{1\over 3},\nonumber
\end{equation}
with $+$ $(-)$ for $\tau =$ neutron (proton). The harmonic
oscillation parameter is given by $b^2_\tau=b^2_0/a_\tau$ with
$b^2_0=\hbar/{(m\omega_0)}=A^{1\over 3}$ fm$^2$. With Eq.
(\ref{chi}) and the deformation parameters in Table
\ref{TableDeforPara}, the QQ-force strength $\chi$ for all nuclei
studied in the present work can then be determined and are shown in
Table \ref{ChiValue}. The monopole pairing strength $G_M$ (in MeV)
is of the standard form
\begin{equation}
G_M = {{G_1 - G_2{{N-Z}\over A}}\over A} ~{\rm for~neutrons,}~~~~
G_M = {G_1 \over A} ~{\rm for~protons.} \label{pairing}
\end{equation}
In the present calculation, we take $G_1=20.82$ and $G_2=13.58$,
which approximately reproduce the observed odd-even mass difference
in the mass region. This choice of $G_M$ is appropriate for the
single-particle space employed in the model, where three major
shells are used for each type of nucleons ($N=3,4,5$ for both
neutrons and protons). The quadrupole pairing strength $G_Q$ is
assumed to be proportional to $G_M$, and the proportionality
constant being fixed as 0.18. These interaction strengths are
consistent with those used earlier for the same mass region
\cite{js01,rp01}.

\begin{figure}[htb]
 \centerline{\includegraphics[trim=0cm 0cm 0cm
0cm,width=0.45\textwidth,clip]{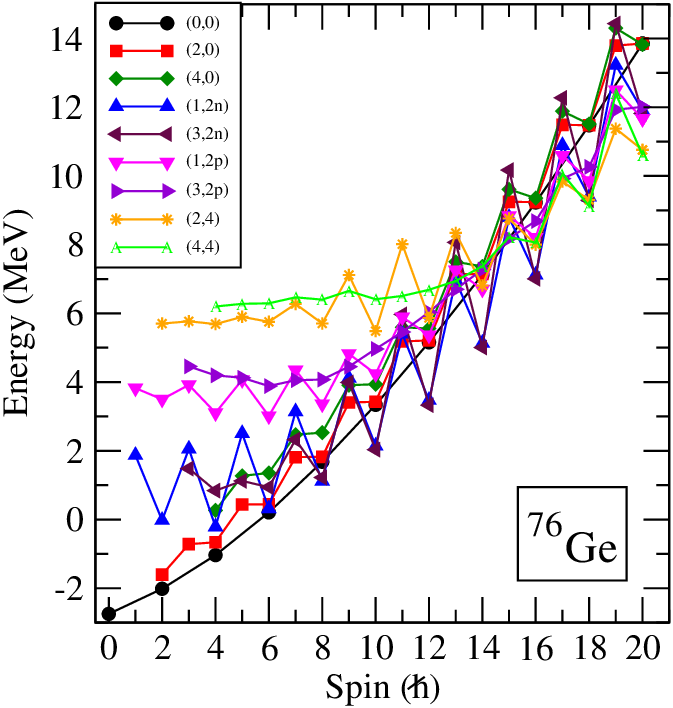}} \caption{(Color
online) Theoretical band diagram for $^{76}$Ge. The labels $(K,\#)$
characterize the states, with $K$ denoting the $K$ quantum number
and $\#$ the number of quasiparticles. For example, (0,0), (2,0),
and (4,0) correspond to the $K=0$ ground-, $K=2$ $\gamma$-, and
$K=4$ $\gamma$$\gamma$-band, respectively, projected from the 0-qp
state. (1,2n), (3,2n), (1,2p), (3,2p), (2,4), and (4,4) correspond,
respectively, to the projected 2-neutron-aligned state,
2-proton-aligned state, 2-neutron-plus-2-proton aligned state, with
different $K$ quantum numbers.} \label{fig1}
\end{figure}
\begin{figure}[htb]
 \centerline{\includegraphics[trim=0cm 0cm 0cm
0cm,width=0.45\textwidth,clip]{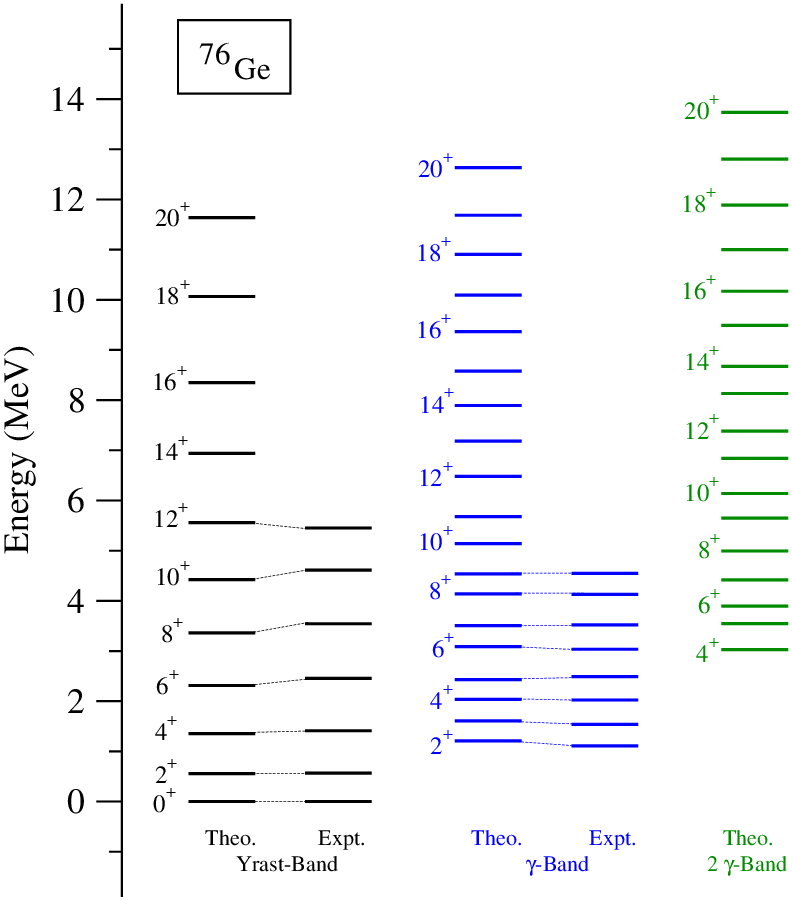}}
 \caption{(Color online) Comparison of the calculated band
energies with available experimental data for $^{76}$Ge. Data are
taken from Ref. \cite{YT13}.} \label{fig2}
\end{figure}

\begin{figure}[htb]
 \centerline{\includegraphics[trim=0cm 0cm 0cm
0cm,width=0.45\textwidth,clip]{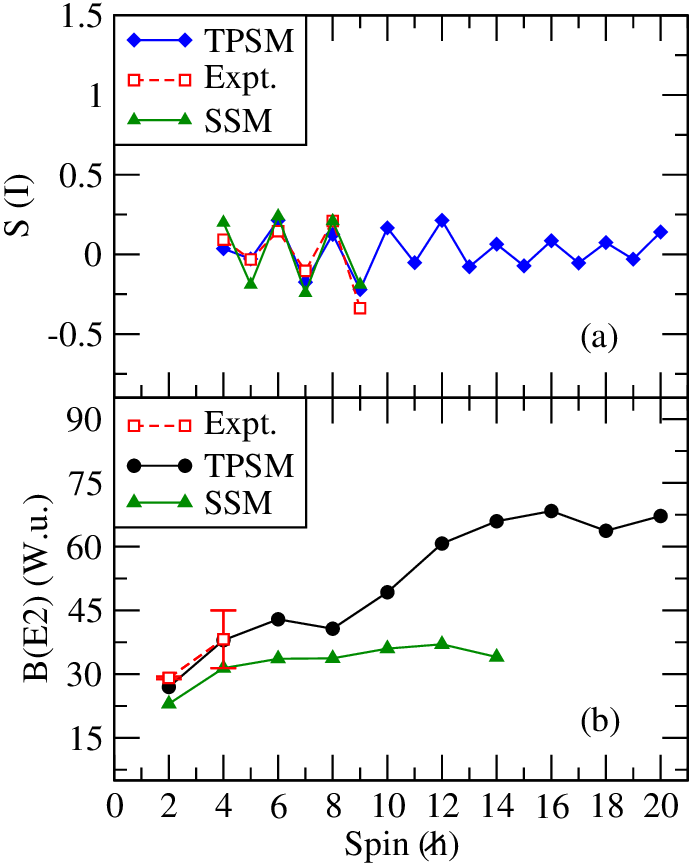}} \caption{(Color
online) Comparison of the TPSM calculation with experimental data
\cite{YT13} for $^{76}$Ge. Results of the spherical shell model
(SSM) calculations are also shown. (a) Staggering parameter S(I) for
the $\gamma$ band, and (b) B(E2) values for the yrast band. The
B(E2) data are taken from Ref. \cite{BE2}. } \label{fig3}
\end{figure}

\begin{figure}[htb]
 \centerline{\includegraphics[trim=0cm 0cm 0cm
     0cm,width=0.45\textwidth,clip]{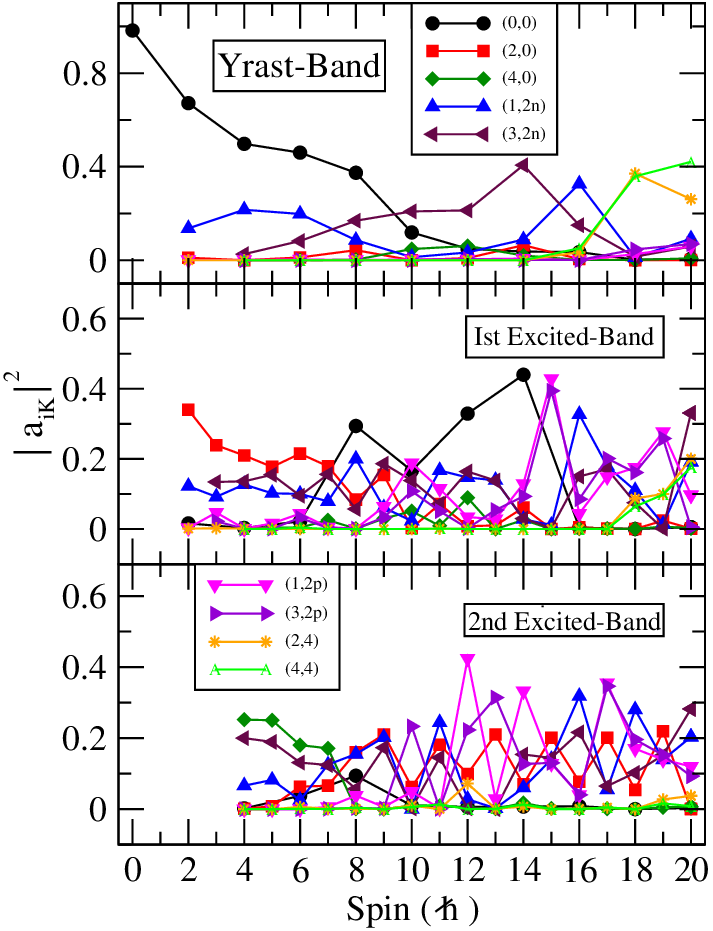}} \caption{(Color online)
Probabilities of the projected configurations in the yrast-, 1st-,
and 2nd-excited bands.} \label{fig4}
\end{figure}
\begin{figure}[htb]
 \centerline{\includegraphics[trim=0cm 0cm 0cm
 0cm,width=0.35\textwidth,clip]{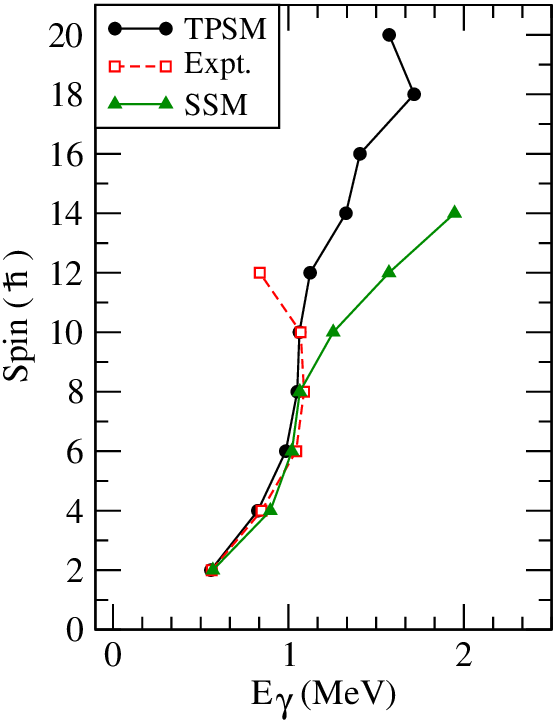}} \caption{(Color online)
Comparison of the TPSM calculation with experimental data
\cite{YT13} for $^{76}$Ge for the relation between spin $I$ and
transition energy $E_\gamma$. Results of the spherical shell model
(SSM) calculations \cite{NY08} are also shown. } \label{fig5}
\end{figure}
\begin{figure}[htb]
 \centerline{\includegraphics[trim=0cm 0cm 0cm
0cm,width=0.45\textwidth,clip]{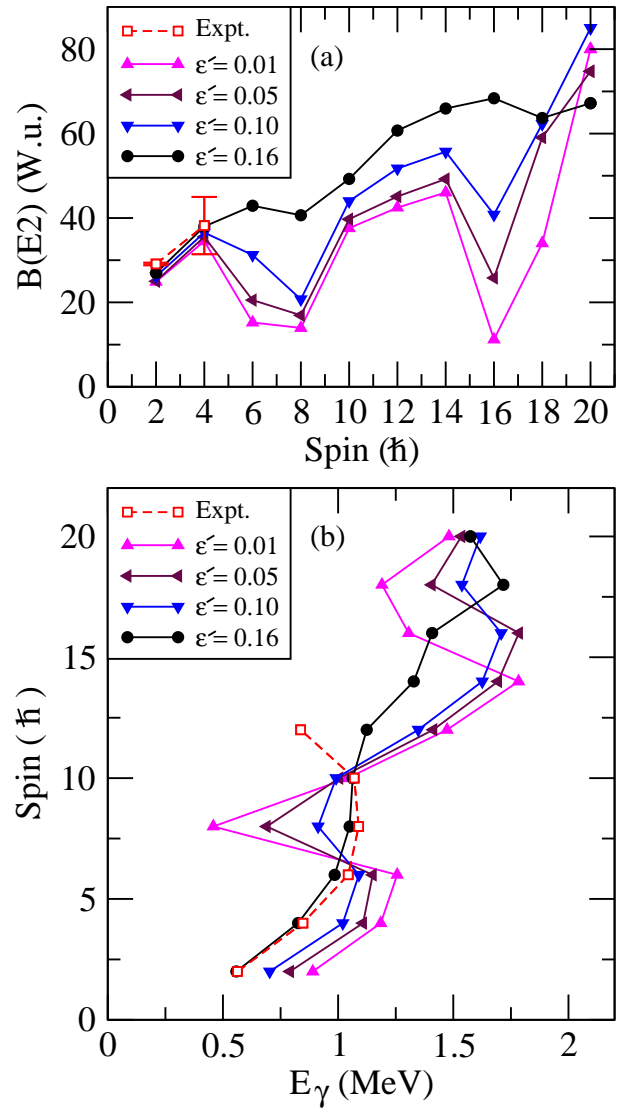}} \caption{(Color
online) (a) Calculated B(E2) values and (b) transition energies
$E_\gamma$ for the yrast-band of $^{76}$Ge with varying
$\epsilon'$.} \label{fig6}
\end{figure}

\section{Results of $^{76}$Ge and rigid $\gamma$-deformation}

TPSM calculations proceed in several stages. In the first stage, the
deformed basis space is constructed by solving the
triaxially-deformed Nilsson potential. In the present work, we have
employed $\epsilon=0.20$ and $\epsilon'=0.16$ (see Table
\ref{TableDeforPara}) in the Nilsson potential to generate the
deformed basis for $^{76}$Ge. The value of $\epsilon$ has been
adopted from the earlier study \cite{LU07} and the value of
$\epsilon'$ has been chosen so that the behavior of the $\gamma$
band is properly described. We shall discuss later the dependence of
the calculation on the triaxial parameter. Pairing is described by
performing a BCS calculation for the single-particle states
generated by the triaxially-deformed Nilsson potential. In the
present work, no particle-number projection is included, and
therefore, this quantum number is conserved only on the average at
the BCS level. In the second step, the good angular-momentum states
are obtained from the deformed basis by employing the
three-dimensional angular-momentum projection technique. The
projected bands obtained from 0-, 2-, and 4-qp states close to the
Fermi surface are displayed in Fig.~\ref{fig1} (the so-called band
diagram, see Ref. \cite{KY95}). The projection from the 0-qp
configuration gives rise to band structures with $K=0,2,4$,
corresponding to the ground-, $\gamma$- and $\gamma\gamma$-band
\cite{YK00}. The calculated band-head energy of the $\gamma-$ and
$\gamma\gamma$-bands are about 1.21 MeV and 3.03 MeV, respectively,
above the ground state.

It is observed from Fig.~\ref{fig1} that the projected bands from
2-quasineutron state having $K=1$ and 3 cross the ground-state band
at $I=8$. These bands are the $\gamma$-band built on the
2-quasineutron-aligned configurations. The 2-quasiproton states are
at higher excitation energies as compared to the 2-neutron states,
and therefore, do not cross the ground-state band. Further, at
$I=18$, the 4-qp structures (2-quasineutron plus 2-quasiproton)
having $K=2$ and 4 cross the yrast-configuration. We stress that in
Fig.~\ref{fig1}, only the lowest bands are displayed for clarity. In
the actual analysis, we use more than thirty-five configurations in
the mixing for each spin-state.

In the third and the final stage, the projected basis are used to
diagonalize the shell model Hamiltonian, Eq.~(\ref{hamham}). The
band energies, obtained after diagonalization, are shown in
Fig.~\ref{fig2} with the available experimental data. It is evident
from the figure that TPSM results are in excellent agreement with
the known experimental energies. In Fig.~\ref{fig2}, the excitation
spectrum is predicted for the $\gamma\gamma$-band, and we hope that
this well-developed band will be populated in future experimental
studies.

In order to understand the nature of the triaxial shape in
$^{76}$Ge, the staggering parameter, defined as,
\begin{equation}
S(I) = \frac{[E(I)-E(I-1)]-[E(I-1)-E(I-2)]}{E(2^{+}_1)} \label{SI}
\end{equation}
is plotted for the $\gamma$-band in Fig.~\ref{fig3}(a). In the same
figure we also provide the existing results of the SSM approach
\cite{NY08}. It is evident from the figure that the experimental
staggering parameter for the known energy levels is reproduced quite
accurately by the TPSM calculations and also by the SSM study. The
TPSM results indicate that above spin $I=10$, the staggering
amplitudes become smaller, and the reason for this is due to a
considerable mixing of the 2-qp configurations with the
$\gamma$-band at higher spins. In order to probe the mixing, the
probabilities of various projected configurations are plotted in
Fig.~\ref{fig4} for the yrast, the 1st-, and the 2nd-excited bands.
The yrast band up to $I=8$ is dominated by the 0-qp configuration
with $K=0$, and above this spin the 2-neutron-aligned band is the
dominant configuration. Above $I=16$, the yrast band is primarily
composed of 4-qp configurations. The 1st-excited band has the
dominant $K=2$ 0-qp configuration until $I=7$ and, therefore, is the
$\gamma$-band. However, above $I=7$, the 1st-excited band has $K=0$
dominant component. The 2-nd excited band has dominant $K=4$ 0-qp
configuration, referred to as $\gamma\gamma$-band, up to $I=7$.
Above this spin value, mixed structures are obtained. The $K=2$
state from the 0-qp configuration seems to become important along
with some 2-qp configurations.

We have also evaluated quadrupole transition probabilities along the
yrast band in the framework of TPSM \cite{JY01}. The standard
effective charges ($e_\pi=1.5e$ and $e_\nu=0.5e$) are used in the
calculation for $^{76}$Ge, and later for all other nuclei studied in
the present work. Experimentally, data for the lowest two
transitions in the yrast band of $^{76}$Ge are available \cite{BE2}.
In the lower panel of Fig.~\ref{fig3}, the $B(E2)$ transition
probabilities are plotted as a function of spin. The calculated
transitions from the SSM approach \cite{NY08} are also displayed in
the figure for comparison. It is seen from the figure that the TPSM
results reproduce the lowest two known transitions quite well while
the SSM values \cite{NY08} are somewhat under-predicted. The
calculated transitions using the TPSM approach predict a drop at
$I=8$ due to the crossing of the 2-quasineutron-aligned band at this
spin value. Above $I=8$, the $B(E2)$ transitions are predicted to
increase rapidly with spin and then drop again at $I=18$ due to the
alignment of two more quasiprotons. On the other hand, the
SSM-predicted transitions depict an increase for the $I=4
\rightarrow$ 2 transition, but above this spin value the SSM
transitions show almost a constant behavior. Thus, there appears to
be a discrepancy between the TPSM and SSM results for the transition
probabilities and it is highly desirable to perform the lifetime
measurements for the high-spin states in $^{76}$Ge. As quadrupole
transition probabilities measure the collective behavior of a
system, a correct description of it usually requires a sufficiently
large model space \cite{YS09}.

\begin{table}
\caption{Ratios of B(E2) rates between states with initial
spin-parity $I^\pi_i$ and final $I^\pi_{f1}$ and $I^\pi_{f2}$, given
by R = B(E2; $I^\pi_i$ $\rightarrow$ $I^\pi_{f1}$)/ B(E2; $I^\pi_i$
$\rightarrow$ $I^\pi_{f2}$). Experimental values \cite{YT13} are
compared with those calculated by the TPSM, the Davydov and Filippov
model (DF), and the spherical shell model (SSM) \cite{NY08}.}
\begin{tabular}{ccccccc}\hline
\hline $I^\pi_{i}$ & $I^\pi_{f1}$ & $I^\pi_{f2}$ &R$_{\rm{Expt.}}$ &R$_{\rm{TPSM}}$ &R$_{\rm{DF}}$ &R$_{\rm{SSM}}$\\
\hline
$2_2^+$ & $0_1^+$ & $2_1^+$  &  0.027 (\textit{3)}  &  0.05  &  0  &  0.04 \\
$3_1^+$ & $2_1^+$ & $2_2^+$  &  0.029(\textit{$^{+6}_{-4}$)}   & 0.04   &  0  &  0.06 \\
$4_2^+$ & $4_1^+$ & $2_2^+$  &  1.34(\textit{4)}   & 1.62  &  0.46  &  0.93 \\
$5_1^+$ & $4_2^+$ & $3_1^+$  &  $<$6.3  &1.35 &  1.0 &  1.29 \\
$6_2^+$ & $4_1^+$ & $4_2^+$  &  0.038(\textit{14)}   &0.19 &  0 &
0.48 \\\hline\hline
\end{tabular}\label{TableBE2}
\end{table}

In Table \ref{TableBE2}, a comparison is provided for the measured
ratios of the $B(E2)$ transition strengths with TPSM predictions and
also with results obtained using the SSM and DF model approaches
\cite{YT13}. It is noted that both TPSM and SSM provide a reasonable
description of the known transitions.

\begin{figure}[htb]
 \centerline{\includegraphics[trim=0cm 0cm 0cm
0cm,width=0.47\textwidth,clip]{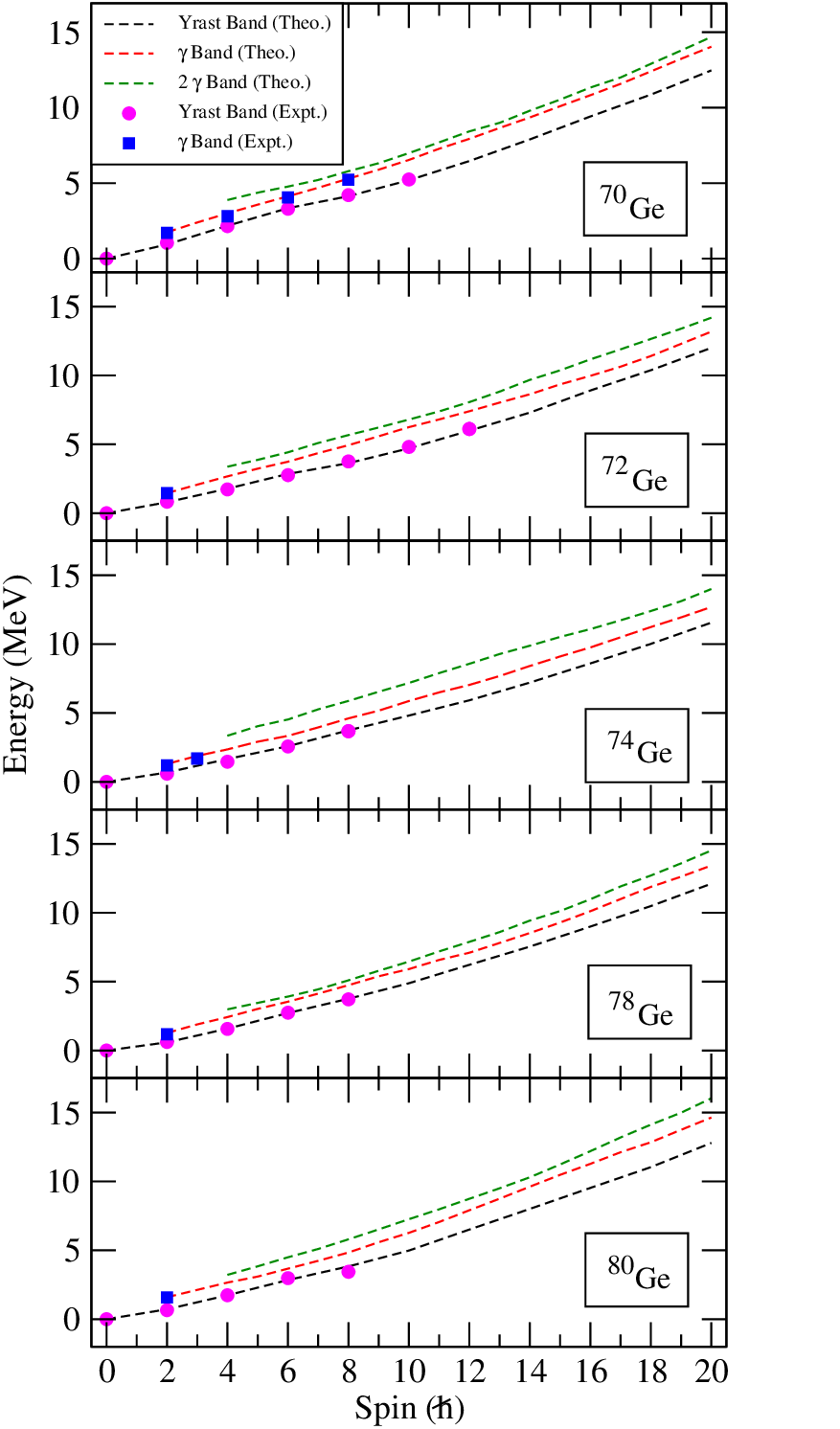}}
 \caption{(Color online) Comparison of the calculated band
energies with available experimental data for $^{70,72,74,78,80}$Ge.
Data are taken from Ref. \cite{DA70,DA72,DA74,DA78,DA80}.} \label{fig7}
\end{figure}
\begin{figure}[htb]
 \centerline{\includegraphics[trim=0cm 0cm 0cm
0cm,width=0.47\textwidth,clip]{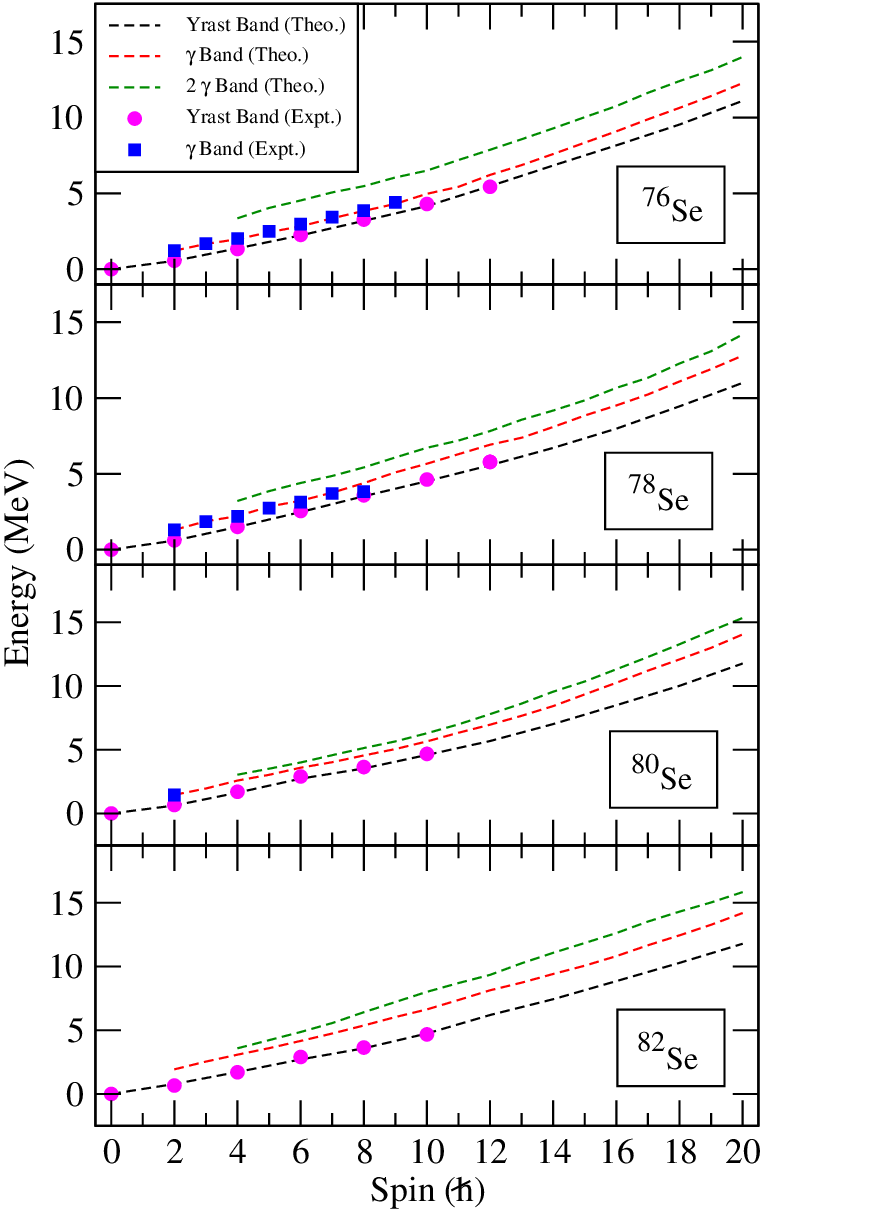}}
 \caption{(Color online) Comparison of the calculated band
energies with available experimental data for $^{76-82}$Se. Data are
taken from Ref. \cite{DA78,DA80,DA76,DA82}.} \label{fig8}
\end{figure}

Aligned quasiparticles carry valuable information about the single
particle structures in the neutron-rich mass region. To explore the
alignment behavior in $^{76}$Ge, angular-momentum is displayed in
Fig.~\ref{fig5} as a function of transition energy $E_{\gamma}$ for
the measured data, which is compared with the present TPSM results
and the corresponding SSM ones. It is clearly seen that the three
curves coincide with each other at low-spins, indicating an
excellent agreement of both the calculations with experiment.
However, it is noted that after $I=8$, the SSM results deviate from
the experimental ones for higher $E_{\gamma}$. The TPSM results, on
the other hand, appear to give a better description of the data,
although, it also cannot reproduce the data point at $I=12$. For
high spin states, TPSM predicts smaller $E_{\gamma}$, thus larger
moments of inertia for this nucleus. The predicted TPSM behavior can
be understood as the results of mixing of multi-qp states at high
spins (see Fig. \ref{fig4} and discussions), which continuously
supplies angular momentum to the system as spin increases. There
could be several reasons for the discrepancy noted at I=12 in
Fig.~\ref{fig5}. We consider the major reason could be due to
constant pairing approximation used in the present TPSM approach.
The BCS equations are solved for the ground-state and same pairing
solution obtained is employed for all the states. This is clearly a
crude approximation for high-spin state as it is known that pairing
correlations are reduced for these states.

In order to investigate the importance of the triaxiality on the
high-spin properties in $^{76}$Ge, the spin-dependence of $B(E2)$
transition probabilities and the transition energies are plotted in
Fig.~\ref{fig6} for varying values of $\epsilon'$. In the upper
panel, for all values of $\epsilon'$, $B(E2)$ show drops at about
$I=8$ and 16 corresponding to band-mixings. However, for lower
values of $\epsilon'$, substantial drops indicate more sudden
changes in the wave functions as compared to the case of
$\epsilon'=0.16$. The angular-momentum plot against $E_{\gamma}$ in
the lower panel of Fig.~\ref{fig6} depicts sharp backbends for lower
values of $\epsilon'$, again due to sharper band-crossings. For
higher values of $\epsilon'$, angular-momentum plot shows a smooth
upward trend and for $\epsilon'=0.16$ the behavior agrees with the
experimental data, corresponding to the triaxiality parameter
$\gamma \approx 30^\circ$.

We would like to add that successful application of the DF model for
$^{76}$Ge to describe the observed $\gamma$ band \cite{YT13} favors
the picture of a rigid-$\gamma$ deformation for this system.
Nevertheless, this model is clearly an over-simplified approach. It
has been pointed out \cite{YK00,YS08} that the underlying physical
picture of generating $\gamma$-vibration in deformed nuclei,
suggested in the framework of TPSM, is analogous to the classical
picture of Davydov and Filippov \cite{AS58}, yet TPSM is a fully
microscopic method. It is interesting to see that both shell models
(SSM and TPSM), though starting from quite different bases
(spherically symmetric vs. triaxially deformed) give nearly
identical results for the low-lying states of $^{76}$Ge, as seen in
Figs. \ref{fig3} and \ref{fig5}, as well as Table \ref{TableBE2}.
Deviations of the results of TPSM from SSM are predicted for
high-spin states (Fig. \ref{fig5}). The extension of measurements to
higher spin is highly desirable as this will shed light on the
limitations of the SSM and the TPSM approaches.

\section{Results and discussions of the neighboring nuclei}

It has been pointed out in ref. \cite{YT13} that $^{76}$Ge is a
unique example in this mass region that depicts a rigid
$\gamma$-deformation with the staggering phase of the $\gamma$-band
in conformity with the DF model. All other nuclei in the
neighborhood have staggering phase opposite to that $^{76}$Ge and
are categorized as $\gamma$-soft nuclei. It is, therefore, quite
interesting to study neighboring nuclei, as well, in order to probe
the mechanisms behind the opposing staggering phase of $^{76}$Ge in
relation to its neighbors. To have a complete overview for the mass
region, we have performed extensive calculations for other even-even
Ge-isotopes, $^{70,72,74,78,80}$Ge, as well as for some Se-
isotopes, $^{76,78,80,82}$Se. For these calculations, the axial
deformations $\epsilon$ are taken from Ref. \cite{Raman} (converted
from $\beta$ to $\epsilon$ by multiplying by 0.95 factor) and the
values are listed in Table \ref{TableDeforPara}. The values for
$\epsilon'$, given also in Table \ref{TableDeforPara}, are chosen in
such a way that the observed band head of the $\gamma-$band is
reproduced. In a few cases where the $\gamma$-band has not been
observed, the $\epsilon'$ of the neighboring nucleus is adopted. The
interaction strengths in Eqs. (\ref{hamham}) and (\ref{pairing}) are
kept the same as in the $^{76}$Ge calculation. In Figs. \ref{fig7}
and \ref{fig8}, the calculated band energies for these nuclei are
compared with the available experimental data. The results clearly
indicate that TPSM approach also provides a good description for
these nuclei apart from $^{76}$Ge.

\begin{figure}[htb]
 \centerline{\includegraphics[trim=0cm 0cm 0cm
0cm,width=0.45\textwidth,clip]{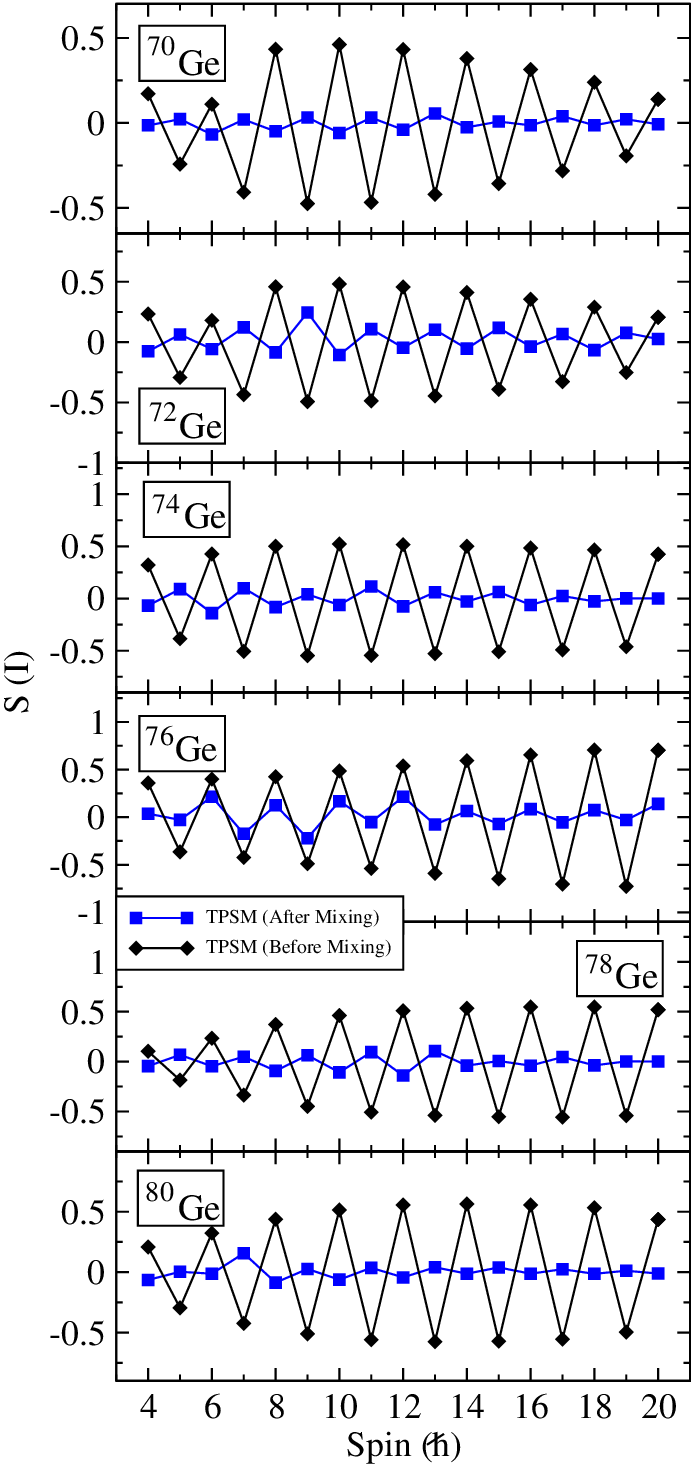}} \caption{(Color
online) Comparison of calculated staggering parameter S(I) for the
$\gamma$ band before and after configuration mixing for
$^{70-80}$Ge. S(I) parameter before mixing are divided by a factor of three
so that they fit in the figure.} \label{fig9}
\end{figure}
\begin{figure}[htb]
 \centerline{\includegraphics[trim=0cm 0cm 0cm
0cm,width=0.45\textwidth,clip]{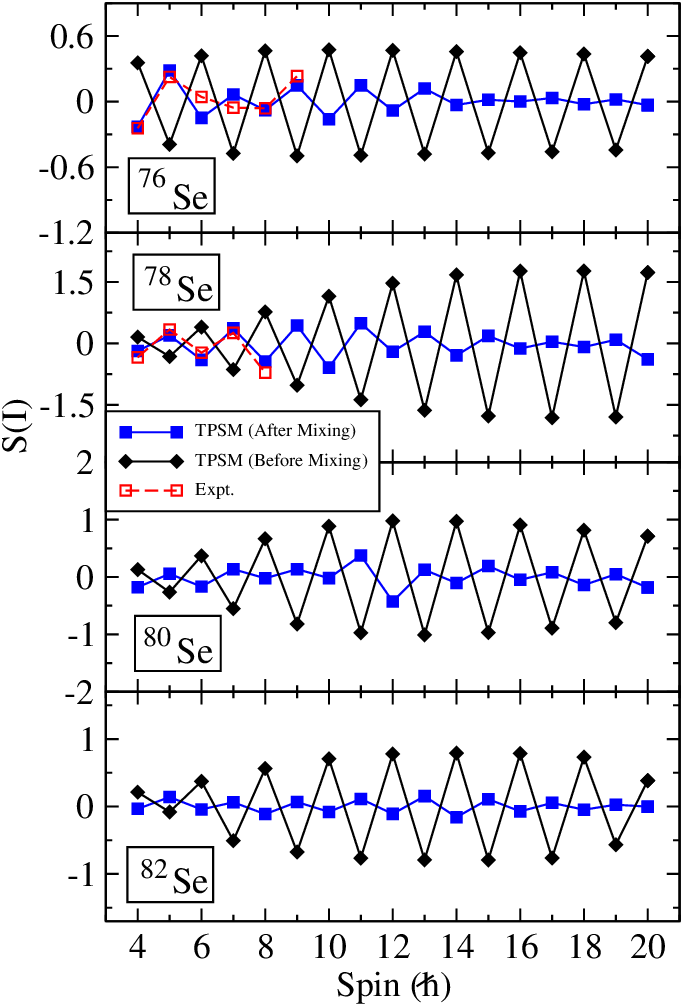}} \caption{(Color
online) Comparison of calculated staggering parameter S(I) for the
$\gamma$ band before and after configuration mixing for
$^{76-82}$Se. S(I) parameter before mixing are divided by a factor of three
so that they fit in the figure. } \label{fig10}
\end{figure}

We shall now turn to the discussion of the staggering phase of the
nuclei depicted in Figs. \ref{fig7} and \ref{fig8} in relation to
$^{76}$Ge. First of all we would like to mention that the model
space in TPSM is constructed from a triaxially-deformed basis with a
given set of deformation parameters $(\epsilon, \epsilon')$ shown in
Table \ref{TableDeforPara}. There are no explicit phonon or
vibrational degrees of freedom in the model. Naively, a model based
on a fixed triaxial deformation is of the kind of Davydov and
Filippov model \cite{AS58}. However, the TPSM is a fully microscopic
theory, and fixed deformations are only used for construction of
basis states. It is important to note that unlike the
phenomenological asymmetric rotor model \cite{AS58}, our results
depend not only on the deformation parameters but also
on the detailed microscopic isotope-dependent shell filling, and
more importantly, on the configuration mixing of the various
quasiparticle states \cite{YK00,YS08}. We would also like to remind
here that in the spherical shell model approach, although, starting
from a bare spherical basis, it can equally describe the
deformed nuclei as well.

The theoretical results of staggering parameter $S(I)$ (see Eq.
(\ref{SI})) for Ge- and Se-isotopes are plotted in Figures
\ref{fig9} and \ref{fig10} before and after mixing of
configurations. What is plotted in Figs. \ref{fig9} and \ref{fig10}
are the full TPSM results after mixing, as shown in Figs. \ref{fig7}
and \ref{fig8}, and those for the projected 0-qp state with $K=2$
[labeled in Fig. \ref{fig1} as (2,0)] only. The latter represents
the major component of the $\gamma$ band \cite{YK00}. The comparison
is made systematically for the Ge and Se isotopes, and therefore,
one may see the effect of isotope-dependent shell-filling.

It is noted from Figs. \ref{fig9} and \ref{fig10} that before
configuration mixing, the calculated $S(I)$ (in black diamonds) show
a rather similar spin-dependent behavior for all the ten nuclei
under consideration. In particular, all of them have the same
staggering phase in $S(I)$. However, the results turn out to be
extremely interesting after a full mixing of quasiparticle
configurations shown in Eq. (\ref{basis}). After the configuration
mixing, only the staggering phase of the $S(I)$ (in blue squares)
for $^{76}$Ge remains unchanged while all other nuclei depict an
opposite phase as compared to $^{76}$Ge. We may thus conclude that
the staggering pattern of $S(I)$ is determined by the configuration
mixing, which is isotope-dependent. A strong mixing of the
configurations in the TPSM basis (\ref{basis}) can lead to
modifications in the nuclear shape, as shown in Figs. \ref{fig9} and
\ref{fig10}, from a rigid triaxial rotor to the one that is soft in
$\gamma$ deformation when interpreted in terms of two extreme
phenomenological models of $\gamma$-rigid of Davydov and Flippov and
$\gamma$-soft of Wilets and Jean.

\begin{figure}[htb]
 \centerline{\includegraphics[trim=0cm 0cm 0cm
0cm,width=0.45\textwidth,clip]{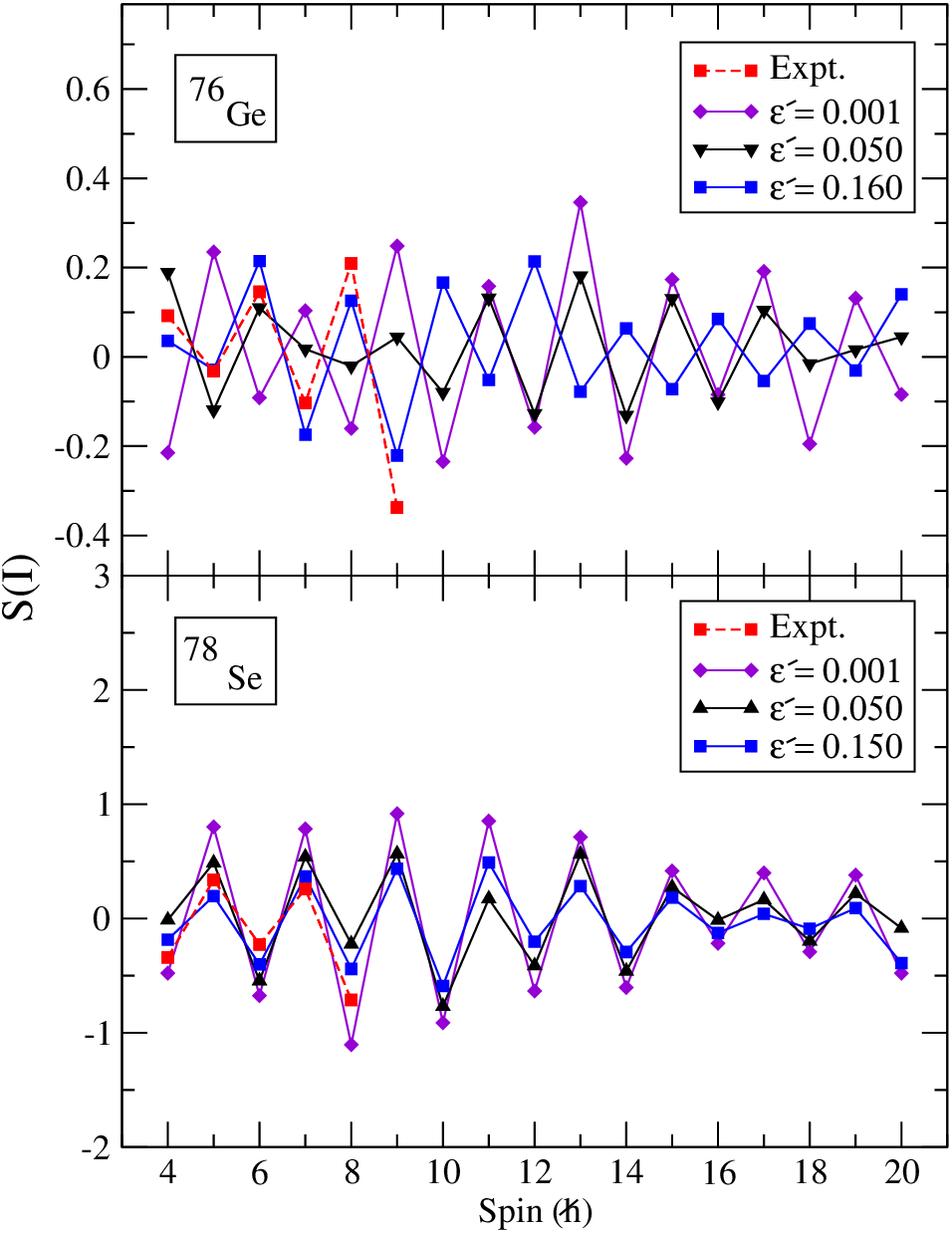}}
\caption{(Color online) Comparison of calculated staggering
parameter S(I) for the $\gamma$ band (results after configuration
mixing) with different triaxial deformation parameters $\epsilon'$
for $^{76}$Ge and $^{78}$Se. } \label{fig10'}
\end{figure}

In order to gain further insight on the above results, we have
calculated the staggering parameter $S(I)$ as a function of
$\epsilon'$, with the results displayed in Fig. \ref{fig10'}. These
results are obtained after the configuration mixing with varying
triaxial deformation in the Nilsson Hamiltonian that generate the
intrinsic basis. It is seen that for $^{76}$Ge, the
experimentally-observed phase of the staggering is reproduced only
for a large value of $\epsilon'$. In contrast, for all other
isotopes the phase is independent of $\epsilon'$, with $^{78}$Se as
an illustrative example in Fig. \ref{fig10'}.

\begin{figure}[htb]
 \centerline{\includegraphics[trim=0cm 0cm 0cm
0cm,width=0.45\textwidth,clip]{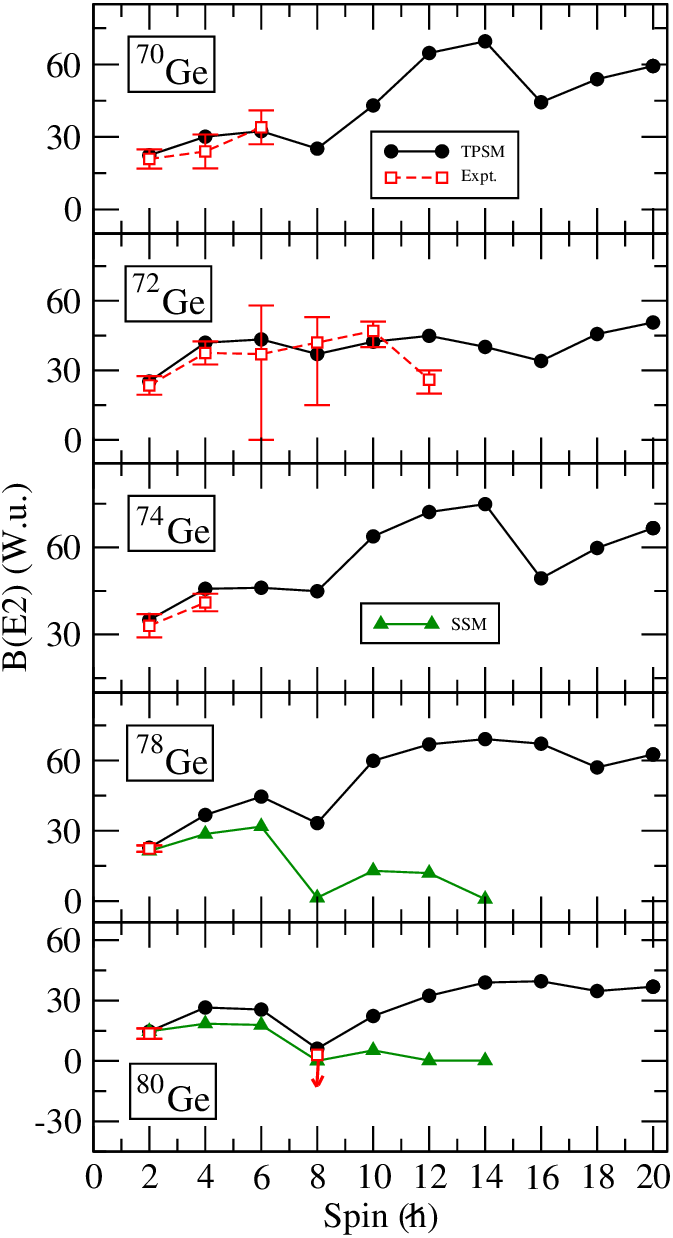}} \caption{(Color
online) Comparison of the TPSM calculation of B(E2) values for the
yrast band with experimental data
\cite{DA70,DA72,DA74,DA78,DA80,TH78,RS78,AM807,AM802} for
$^{70-80}$Ge. Results of the spherical shell model (SSM)
calculations \cite{NY08} are also shown for the available nuclei. }
\label{fig11}
\end{figure}
\begin{figure}[htb]
 \centerline{\includegraphics[trim=0cm 0cm 0cm
0cm,width=0.45\textwidth,clip]{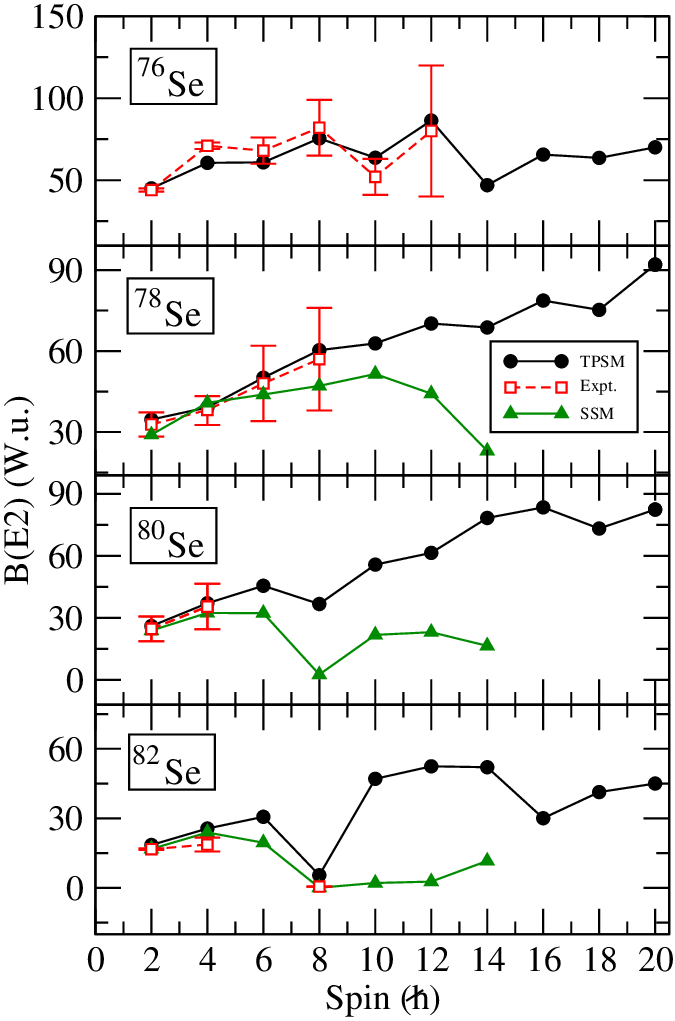}} \caption{(Color
online) Comparison of the TPSM calculation of B(E2) values for the
yrast band with experimental data \cite{DA78,DA80,DA82,DA76,76RL,TH78,RS78,AM807,AM802,KH820} for $^{76-82}$Se.
Results of the spherical shell model (SSM) calculations \cite{NY08}
are also shown for the available nuclei. } \label{fig12}
\end{figure}

We have also calculated the $B(E2)$ values along the yrast band for
$^{70,72,74,78,80}$Ge and $^{76,78,80,82}$Se, and compared them with
available experimental data in Figs. \ref{fig11} and \ref{fig12}.
The calculated $B(E2)$s from the SSM approach \cite{NY08} are also
displayed in the figures for comparison. As is evident from these
figures, the TPSM calculations describe the known experimental
$B(E2)$ values quite nicely. The SSM calculation \cite{NY08} for
$^{78,80}$Ge and $^{78,80,82}$Se, although, reproduce well the
existing experimental data, however, as in the $^{76}$Ge case, the
SSM transitions depict an increase for low spins, but drop
significantly at high spins. In particular, above $I=8$ the SSM
transitions show a completely different behavior as compared to the
TPSM calculation which, in general shows an increasing trend toward
higher spins. There appears to be a major discrepancy between the
TPSM and SSM results for the transition probabilities in high-spin
states for all the nuclei studied in the present work.

\begin{table}
\caption{Calculated inter-band $B(E2)$ values (in W.u.) from
$\gamma$ band to ground band for $^{76}$Ge and $^{78}$Se.}
\begin{tabular}{ccc}\hline
\hline $(I,K)_i\rightarrow (I,K)_f$ & $^{76}$Ge & $^{78}$Se\\
\hline
$(2,2)\rightarrow (0,0)$ & 5.39  & 3.59 \\
$(4,2)\rightarrow (2,0)$   & 5.78  & 0.70 \\
$(6,2)\rightarrow (4,0)$   & 4.55  & 1.84\\
 $(8,2)\rightarrow (6,0)$  & 13.38 & 36.54 \\
$(10,2)\rightarrow (8,0)$  & 8.60  & 5.04\\
$(12,2)\rightarrow (10,0)$ & 1.66  & 0.26 \\
$(14,2)\rightarrow (12,0)$ & 0.21  & 0.07 \\
$(16,2)\rightarrow (14,0)$ & 0.09  & 0.34 \\
$(18,2)\rightarrow (16,0)$ & 5.76  & 0.50 \\
$(20,2)\rightarrow (18,0)$ & 2.35  & 0.65 \\
\hline
  $(2,2)\rightarrow (2,0)$ & 33.26 &26.29  \\
  $(3,2)\rightarrow (2,0)$ & 9.15  & 6.14  \\
  $(3,2)\rightarrow (4,0)$   &0.23  &  0.51  \\
  $(4,2)\rightarrow (4,0)$ & 20.90 & 16.76 \\
 $(5,2)\rightarrow (4,0)$  & 9.27  & 3.13 \\
  $(5,2)\rightarrow (6,0)$   &7.05  &  5.74  \\
  $(6,2)\rightarrow (6,0)$ & 10.41 & 9.24 \\
 $(7,2)\rightarrow (6,0)$  & 7.65  & 2.52  \\
  $(7,2)\rightarrow (8,0)$   &6.84  &  7.62 \\
  $(8,2)\rightarrow (8,0)$ & 7.87 & 4.82 \\
$(9,2)\rightarrow (8,0)$   & 6.50  & 9.48\\
  $(9,2)\rightarrow (10,0)$  &3.47  &  6.56 \\
  $(10,2)\rightarrow (10,0)$ &4.14  &7.84  \\
$(11,2)\rightarrow (10,0)$ & 5.13  & 9.67  \\
   $(11,2)\rightarrow (12,0)$ & 0.11 &  2.39 \\
  $(12,2)\rightarrow (12,0)$ &2.27  &7.71  \\
 $(13,2)\rightarrow (12,0)$& 4.15  & 6.62\\
  $(13,2)\rightarrow (14,0)$ &0.19  &  1.04 \\
  $(14,2)\rightarrow (14,0)$ &0.29  & 4.91 \\
 $(15,2)\rightarrow (14,0)$& 0.09  & 4.96\\
  $(15,2)\rightarrow (16,0)$ &0.81  &  0.69 \\
  $(16,2)\rightarrow (16,0)$ &0.36  &2.56  \\
 $(17,2)\rightarrow (16,0)$& 1.25 & 4.86\\
  $(17,2)\rightarrow (18,0)$ &1.73  &  1.25 \\
  $(18,2)\rightarrow (18,0)$ &0.19  & 0.94 \\
 $(19,2)\rightarrow (18,0)$& 1.88 & 5.30\\
  $(20,2)\rightarrow (18,0)$ &2.57  & 1.24 \\
\hline \hline
\end{tabular}\label{LinkingBE2}
\end{table}

In Table \ref{LinkingBE2}, we present the calculated inter-band
B(E2) values that link the $\gamma$ band to the ground band. An
early example of a similar TPSM calculation can be found in Ref.
\cite{Pla02}. We give all the possible linking transitions for the
low-lying states in $^{76}$Ge, together with those for $^{78}$Se as
an illustrative example. It will be quite interesting to compare
these values with the results from other models, for instance, the
O(6) limit of the Interacting Boson Model \cite{CB85,SOG89}.

\begin{figure}[htb]
 \centerline{\includegraphics[trim=0cm 0cm 0cm
 0cm,width=0.45\textwidth,clip]{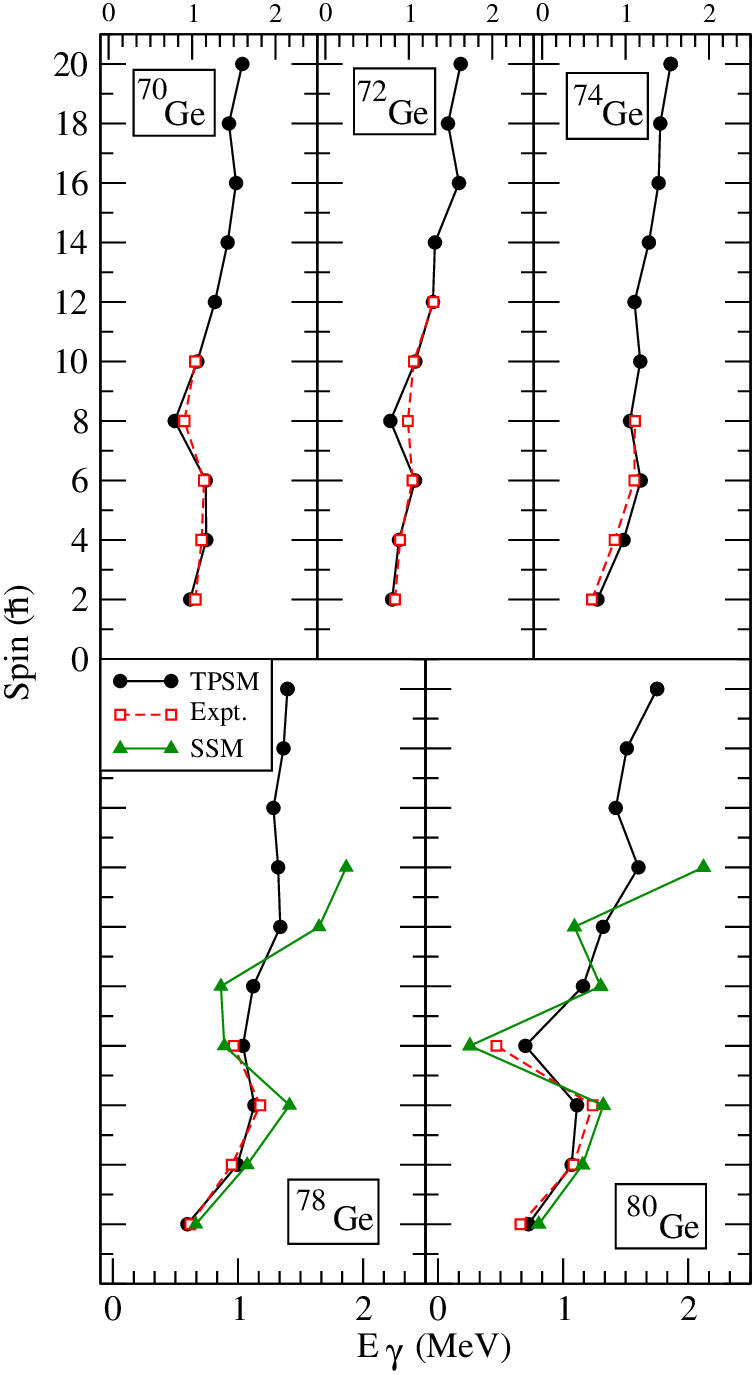}} \caption{(Color online)
Comparison of the TPSM calculation with experimental data \cite{DA70,DA72,DA74,DA78,DA80}
for $^{70-80}$Ge for the relation between spin $I$ and transition
energy $E_\gamma$. Results of the spherical shell model (SSM)
calculations \cite{NY08} are also shown for the available nuclei. }
\label{fig13}
\end{figure}
\begin{figure}[htb]
 \centerline{\includegraphics[trim=0cm 0cm 0cm
 0cm,width=0.45\textwidth,clip]{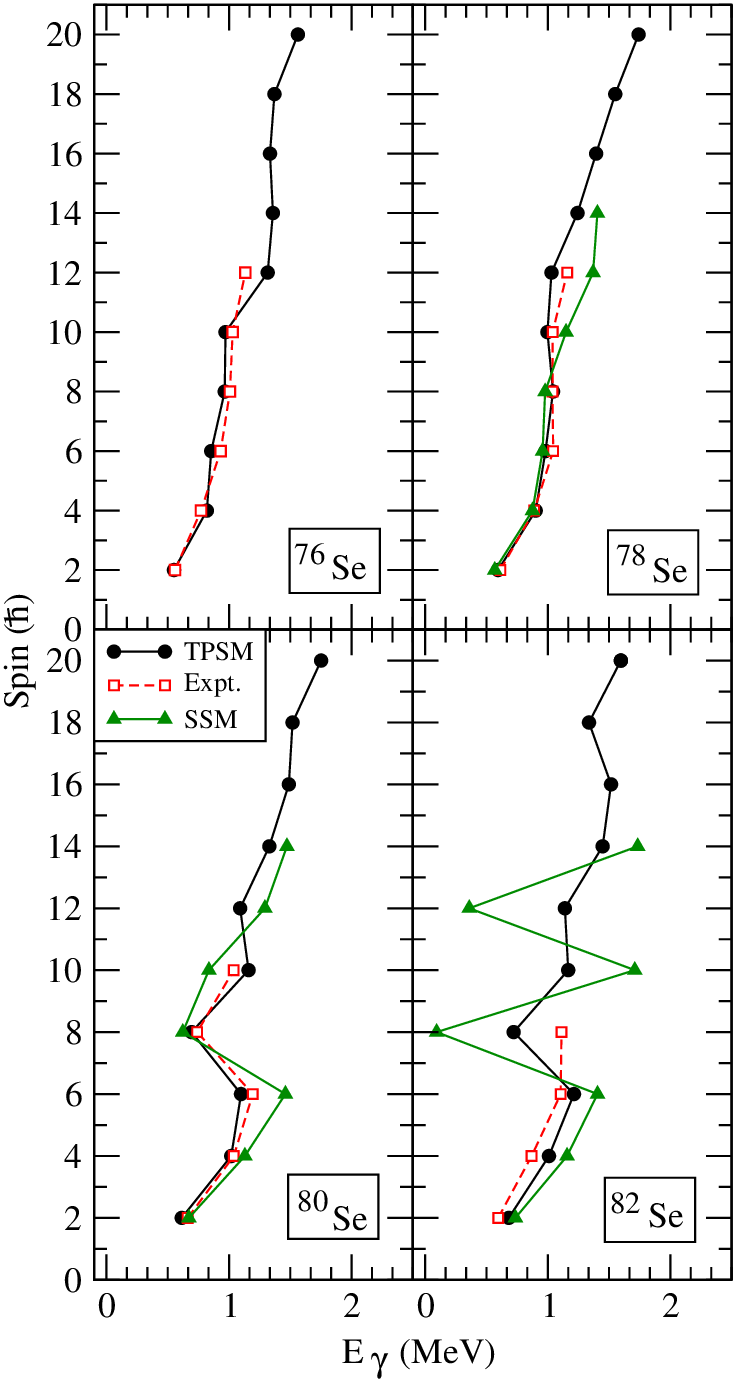}} \caption{(Color online)
Comparison of the TPSM calculation with experimental data \cite{DA78,DA80,DA82,DA76}
for $^{70-80}$Ge for the relation between spin $I$ and transition
energy $E_\gamma$. Results of the spherical shell model (SSM)
calculations \cite{NY08} are also shown for the available nuclei. }
\label{fig14}
\end{figure}

Finally, in Figs. \ref{fig13} and \ref{fig14}, experimentally-known
angular-momenta are displayed as functions of transition energy
$E_\gamma$ for $^{70,72,74,78,80}$Ge and $^{76,78,80,82}$Se, which
are compared with the present TPSM results and the corresponding SSM
ones \cite{NY08}. It is clearly seen that both theoretical
calculations describe the known data very well. Nevertheless, it is
observed, as in the case of $^{76}$Ge, discussed earlier that roughly above
$I = 8$, the TPSM and SSM results deviate from each other for higher
spin states. The predicted SSM values show pronounced zigzag pattern in the
curves while the TPSM results appear more smoother.

\section{Summary}

To summarize, the recently reported experimental measurement for
$^{76}$Ge \cite{YT13} suggested that this nucleus may be a rare
example of a nucleus exhibiting a rigid $\gamma$ deformation in its
low-lying states. Our microscopic calculations using the
multi-quasiparticle triaxial projected shell model support this
inference. By studying various physical quantities, it is shown that
in order to describe the data accurately for both the yrast and
$\gamma$-vibrational bands in $^{76}$Ge, a fixed triaxial
deformation parameter $\gamma\approx 30^\circ$ is required for the
TPSM calculation, which is consistent with that of the DF model
\cite{YT13}. The TPSM results are discussed closely with the
experimental observations and also compared with the previous
spherical shell model calculations \cite{NY08}. Furthermore,
experimental identification of the $\gamma\gamma$- band, predicted
in the present work for this $\gamma$-rigid nucleus, would be very
interesting.

To further demonstrate that the TPSM model with the same parameters
as those of $^{76}$Ge is also applicable to the neighboring nuclei,
we have made a systematic investigation for $^{70,72,74,78,80}$Ge
and $^{76,78,80,82}$Se, and discussed the results. It has been
demonstrated that configuration mixing of various quasiparticle
states can result in a dynamical change for a nucleus from being a
$\gamma$-rigid like to a $\gamma$-soft type when interpreted in
terms of the two phenomenological models of $\gamma$-rigid of
Davydov and Flippov and $\gamma$-soft of Wilets and Jean. The
odd-even staggering phase of the $\gamma$-band is quite opposite in
these two models and has been proposed to be an indicator of the
nature of the $\gamma$-deformation. What we have shown using the
microscopic TPSM model is that the configuration mixing can lead to a
transition from $\gamma$-rigid to $\gamma$-soft phases, at least,
for nuclei studied in the present work. It remains to be explored
whether similar observation is valid for other regions as well.

The $^{76}$Ge nucleus belongs to the group of a few candidates where
neutrinoless double-$\beta$ decay may be observed. In this context,
we note that the recent beyond-mean-field calculations of nuclear
matrix elements for neutrinoless double-$\beta$ decay, based on the
energy density functional method using the Gogny D1S functional,
assumed axial symmetry for the $^{76}$Ge shape \cite{RM10}. As the
nuclear matrix elements serve an important link between
$\beta$-decay observations and the neutrino mass \cite{EV02}, it
remains to be demonstrated what modifications triaxial mean-field
deformation will make in the evaluation of the nuclear
matrix-elements.

\section{Acknowledgement}

Research at Shanghai Jiao Tong University was supported by the
National Natural Science Foundation of China (No. 11135005) and by
the 973 Program of China (No. 2013CB834401).

\end{document}